\documentclass[a4paper,fleqn]{cas-dc}
\usepackage{bm,times}
\usepackage{epstopdf}
\usepackage[sf]{subfigure}
\usepackage[numbers]{natbib}
\ExplSyntaxOn
\cs_gset_eq:NN \msg_term:n \iow_term:n
\ExplSyntaxOff

\begin{document}

\shorttitle{Comments on Solid State Sciences 13 (2011) 251-256 }
\shortauthors{Anand Pal $ et~ al. $ , \today}

\title [mode = title]{Comments on “X-ray analysis of ZnO nanoparticles by Williamson Hall and size-strain plot methods” Solid State Sciences 13 (2011) 251-256}                      

\author[1]{Anand Pal}[orcid=0000-0003-1602-507X]
\cormark[1]
\ead{sandhu.anand@hotmail.com}


\address[1]{Department of Physics, Manipal Institute of Technology Bengaluru,
	Manipal Academy of Higher Education, Manipal, India}

\begin{abstract}
The equation for the size strain plot methods reported by A. Khorsand Zak \textit{et al.}  (Solid State Sci. 13 (2011), 251) \cite{Zak11} does not follow the dimensional homogeneity, consequently leading to an inaccurate estimation of the crystallite size and strain values of the materials under investigation and the dimensions of the obtained parameters. We also perceived an error in the values of crystallite size and strain reported by the authors using the size-strain plot method. We will discuss the importance of dimensional analysis and its repercussions on the estimated values and the units of parameters.
  
\end{abstract}

\maketitle

\section{Introduction}
X-ray diffraction is a versatile and crucial tool for material science researchers. It provides a wide range of information such as unit cell dimensions, crystal symmetry, atomic positions, microstrain, and internal stress of the specimen based on the peak position, intensity, peak profile shape, etc. \cite{Cullity14}. Over the years, many methods have been developed to estimate the crystallite size and strain in the materials using X-ray peak profile analysis. As discussed by A. Khorsand Zak \textit{et al.}\cite{Zak11} the Scherrer formula, Williamson Hall (WH) method, and size strain plot are more common to estimate these parameters from the X-ray peak profile analysis in the materials science community. All these methods have their merits and demerits. However, we restrict our discussion to the equation used by the authors of the aforementioned article for the size strain plot (SSP) method.  

In this short comment, we will discuss the inaccuracy of the equation adopted by the authors for SSP methods and its consequence on the estimated values and the units of the parameters, particularly crystallite size ($D$) and strain ($ \epsilon $) of the specimen. We would suggest the more accurate equation for size strain plot (SSP) methods, which will overcome all the issues of the evaluated parameters' dimensions, units, and numerical values.

\section{Comments}
For the size strain plot method, A. Khorsand Zak \textit{et al.}  \cite{Zak11} used the following equation (Eq. (12) in published paper),
\begin{equation}
	\label{zakssp}
	\left( d_{hkl}\beta_{hkl}\cos\theta\right)^2 = \frac{K}{D}\left( d_{hkl}^2\beta_{hkl}\cos\theta\right) + \left(\frac{\epsilon}{2}\right)^2.
\end{equation}
Where $ d_{hkl} $ is inter-planar spacing, $ K $   is a dimensionless shape factor, with a typical value of 0.9, but varies with the actual shape of the crystallite \cite{Warren90}. $ \lambda $ is the wavelength of the incident X-rays beam, $ \beta_{hkl} $ denotes corrected full width at half maximum, and $ \theta $ is Bragg's angle.

We find multiple issues using Eq.~\eqref{zakssp} for the estimations of crystallite size ($D$) and strain ($ \epsilon $) values. The principal issues and their consequence are discussed below in detail.

\subsection{Inconsistency in the dimensions of Eq.~\eqref{zakssp} }
The X-ray wavelength ($ \lambda $), inter-planar spacing ($ d_{hkl} $), and crystallite size ($ D$) have the dimension of length ($ L $), whereas $ K $, $ \beta_{hkl}$, and $ \theta $  are dimensionless quantities. Using this information, one can confirm that the Scherrer and Williamson Hall equations, Eq. (5) and Eq. (8), respectively,  used by A. Khorsand Zak \textit{et al.} ~\cite{Zak11} are satisfying the dimensional analysis. In both the equations, the dimension of RHS is equal to the dimension of LHS. However, Eq.~\eqref{zakssp} (Eq. (12) in the published article) does not satisfy dimensional analysis. The dimensions of RHS and LHS are not equal.

The dimensions of RHS and LHS for Eq.~\eqref{zakssp}, 
\[ L^2 = \frac{L^2}{L} + L^0\]
\[ \Rightarrow L^2 = L^1 +L^0.\]
Dimensional homogeneity is the basic principle; every equation with physical significance should be dimensionally homogenous. That is to say, the left and right hand sides must have the same dimension for any physically meaningful equation or inequality. If the equation does not follow the dimensional homogeneity, it does not have any physical meaning.
\begin{figure}
	\centering
	\includegraphics[width=0.95\linewidth]{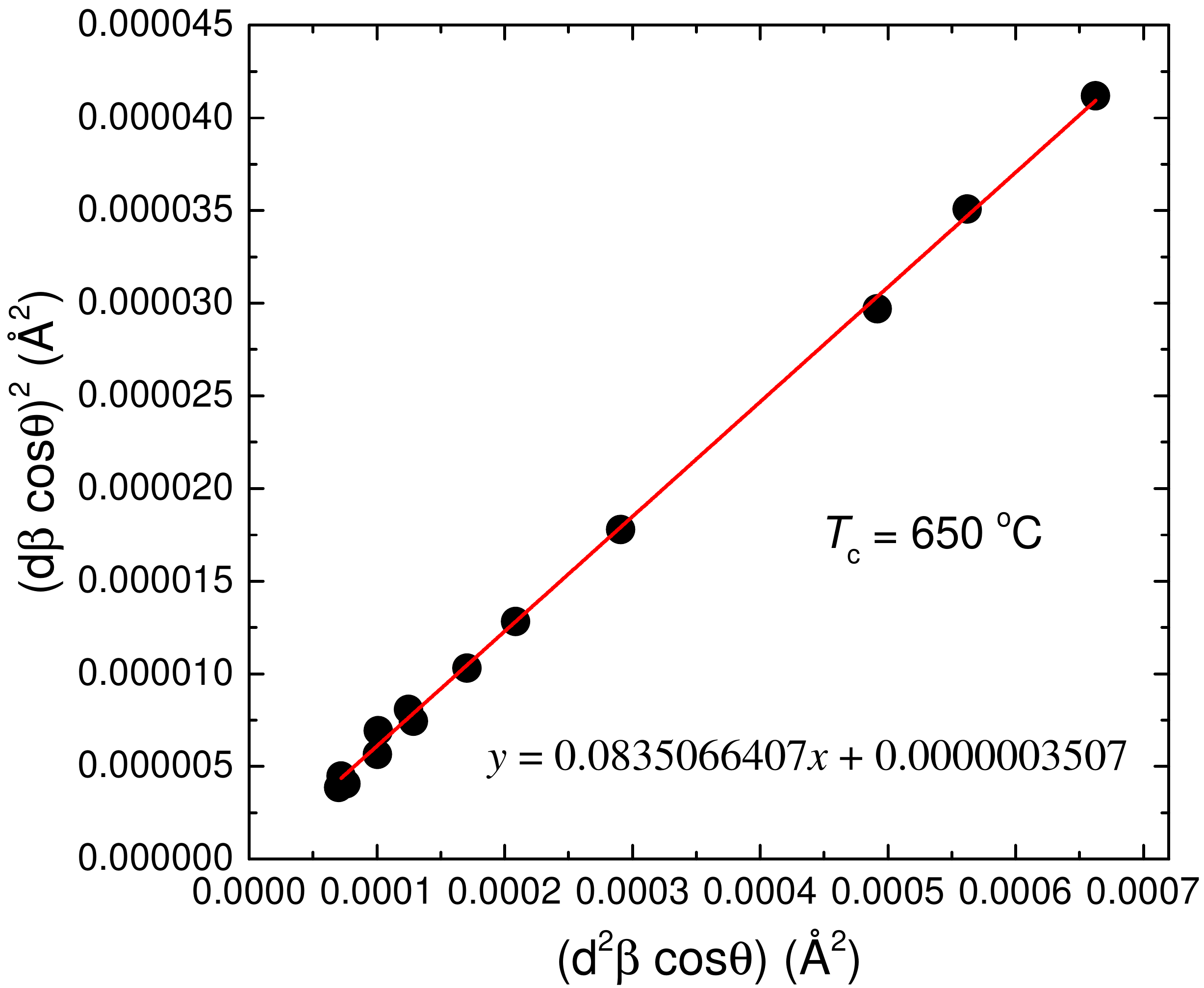}
	\caption{The  size strain plot of ZnO-NPs calcinated at 650~$^o$C. Figure is reproduced from the original paper of   A. Khorsand Zak \textit{et al.}~ \cite{Zak11} }
	\label{fig:ssp-1}
\end{figure}
\subsection{Contradiction  in the units of Crystallite size ($ D $) and Strain ($ \epsilon $) }\label{conta-units}
The interplanar distance can be calculated using the Bragg's law,  $ d_{hkl} = n\lambda/2\sin\theta $, where $ n $ is order of reflection. For simplicity, we consider $ n =1 $.  The X-rays wavelength for Cu-K$ _{\alpha} $  is $ \lambda = 1.5406 $ {\AA}. Then $ d_{hkl} $ is also having the unit of {\AA}. Therefore units of $ x $ and $ y $ axis in Fig. \ref{fig:ssp-1} would be {\AA$ ^2 $} (ignoring the radians for $ \beta_{hkl} $) i.e., both $ x $ and $ y $ axis having the dimension of $L^2$. 

By the definition of slope of the line, $ m = \frac{dy}{dx} $, where $ dy $ and $ dx $ are the  infinitesimal change in $y$-axis with respect to the change in $x$-axis. As both $ x $ and $ y $-axis having the units of {\AA$ ^2 $} (or dimension of $L^2$). The slope of the linearly fitted equation in Fig.~\ref{fig:ssp-1} would be unitless and dimensionless quantity. 

The crystellite size would be,
\[m = \frac{K}{D} \Rightarrow  D = \frac{K}{m}\]

As both $ K $  and $ m $ are dimensionless and unitless quantities, therefore crystalline size ($ D $) would be dimensionless and unitless. It is quite a contradictory result as it is well known that crystalline size has the dimension of length. However, here, we are completely unable to comprehend how the authors of the paper got the unit of the crystallite size ($D$) in nm using Eq.~\eqref{zakssp}, as reported in Table 2 in the published paper. We observe that the authors might have chosen the units of  $D$ arbitrarily without thoroughly inspecting the calculations.  

On the other hand, the intercept ($ C $) of the fitted line would have the unit of  {\AA$ ^2 $} or dimension of $ L^2 $, as it is the value of  $ y $  at $ x \:=\: 0$. If that is the case, it leads to another contradiction for the unit and the dimension of strain ($ \epsilon $).

\[  \left(\frac{\epsilon}{2}\right)^2 = C  \Rightarrow  \epsilon = 2\sqrt{C}\]

As $ C $ is having the unit of {\AA}$ ^2 $, the unit of $ \epsilon $ would be {\AA}. Which is not per the definition of strain. By definition, strain is a ratio of two lengths ($ \epsilon = \Delta L/L$), hence can not have the unit of length and is rather a unit less and dimensionless quantity. The above discussion demonstrates that the SSP equation used by A. Khorsand Zak \textit{et al.}  \cite{Zak11} leads to contradictory units for both crystallite size and strain. 
 \subsection{Incorrect values of Crystallite size ($ D $)}
Using the values of slope and intercept of the linear fit reported in Fig. 7 (a) and (b) by A. Khorsand Zak \textit{et al.}  \cite{Zak11},  we have calculated the values of crystallite size and strain (see Table 1). We are afraid to say that the values of $ D $ and $ \epsilon $ reported by the authors (see Table 2 in ref \cite{Zak11}) are erroneous. It is evident at first glance that the slope values in both the fitting have one order of difference, which directly indicates that there should be a one order of difference in the crystallite size value ($ D $). However, to our surprise, the reported values in Table 2 in the published paper are in the same order.
  
\section{Discussions}
We would like to argue that the Eq.~\eqref{zakssp} is inappropriate, and the correct equation of physical parameters must follow the dimensional homogeneity rule. To resolve the issues of the dimensional homogeneity and units of the crystellite size and strain, an appropriate equation for SSP methods should be modified as \cite{Tatarchuk17,Mangavati22}, 
\begin{equation}
	\label{modifiedssp1}
	\left( \frac{d_{hkl}\beta_{hkl}\cos\theta}{\lambda}\right)^2 = \frac{K}{D}\left( \frac{d_{hkl}^2\beta_{hkl}\cos\theta}{\lambda}\right) + \left(\frac{\epsilon}{2}\right)^2,
\end{equation}
or 
\begin{equation}
	\label{modifiedssp2}
	\left( d_{hkl}\beta_{hkl}\cos\theta\right)^2 = \frac{K\lambda}{D}\left( d_{hkl}^2\beta_{hkl}\cos\theta\right) + \left(\frac{\epsilon \lambda}{2}\right)^2.
\end{equation}

\begin{table*}
	\centering
	\caption{The calculated values of crystallite size ($ D $) and strain ($ \epsilon $) according to Eq.~\eqref{fig:ssp-1} and Eq.~\eqref{modifiedssp1} (or Eq.~\eqref{modifiedssp2}). The subscripts represent the equation number and the corresponding value of shape parameter $ K $, used in the calculation. The values of intercept and the slope are taken from the linear fit reported in Fig. 7 (a) and (b) in the published article  by A. Khorsand Zak \textit{et al.} ~\cite{Zak11} }
	\label{tab:table1}
\begin{tabular}{ccccccccc}
	\hline
	Sample & Slope & Intercept & D$ _{1, K \:= \: 0.75} $   & D$ _{1, K \:= \: 0.9} $ & $ \epsilon_1 $ & D$ _{2, K \:= \: 0.75} $& D$ _{2, K \:= \: 0.9} $ & $ \epsilon_2 $ \\
	&  & ($ \times 10^{-7} $) &  (a.u.) & (a.u.) &  ($ \times 10^{-4} $) & (nm) & (nm) & ($ \times 10^{-4} $) \\
	\hline
	650$^{\circ}$ & 0.08351  & 3.507  & 8.98132 & 10.77759 & 11.8  &  138.3662 & 166.0395 & 7.6879 \\
		\hline
	750$^{\circ}$& 0.75367 & 1.984 & 0.99513  & 1.19416  & 8.90842 & 15.3310 & 18.3972 & 5.7824 \\
	\hline
\end{tabular}
\end{table*}

Both Eq.~\eqref{modifiedssp1} and Eq.~\eqref{modifiedssp2} are dimensionally homogeneous equation.  The dimensions of LHS  and RHS are equal. All the term in Eq.~\eqref{modifiedssp1} and Eq.~\eqref{modifiedssp2} are having the dimensions of  $ L^0 $    and $ L^2 $, respectively. A careful observation of Eq.~\eqref{modifiedssp1} and Eq.~\eqref{modifiedssp2} suggests that it would solve the problem of the contradictory units of $D$ and $ \epsilon $ as discussed in section \ref{conta-units}. We would also like to argue that the Eq.~\eqref{modifiedssp1} and Eq.~\eqref{modifiedssp2} would give a better estimation of the values of crystallite size($ D $) and strain ($ \epsilon $) of the specimen under investigation.

Now, we would like to bring the reader's attention to the repercussions of the  Eq.~\eqref{fig:ssp-1} on the numerical values of  $ D $  and $ \epsilon $. We have calculated the values of $ D $  and $ \epsilon $ using both Eq.~\eqref{fig:ssp-1} and Eq.~\eqref{modifiedssp1} (or Eq.~\eqref{modifiedssp2} ) and exploiting the values slope and the intercept reported by the authors in Fig. 7 (a) and (b).  In the published paper, the value of K is taken 3/4 by the authors. However, in general, the value of K is 0.9. For comparison, we used both the values of $ K $ and the yielded values of $ D $  and $ \epsilon $ are tabulated in Table \ref{tab:table1}. The subscripts shows the equation number ($ D_{1, K \:=\: 0.75} $ means Eq.~\eqref{fig:ssp-1} and K = 0.75 is used for calculation) and the corresponding value of $ K $.

As it is clear from column 4 of Table \ref{tab:table1}, the calculated values of crystallite size using the values of the slope of the linear fit from Fig. 7 (a) and (b) of the published article differ from the values reported by A. Khorsand Zak \textit{et al.}\cite{Zak11} in Table 2. The crystallite size values reported by the authors seem implausible as the slope has one order of difference for both the studied samples; therefore, it is a naïve expectation to have one order of difference in the values of crystallite size ($ D $). However, the values of the $ D $ reported by the authors are nearly identical. Though, the reported values of strain are similar to our calculations. 

We also observed that the estimated values of  crystallite size ($ D $) and  strain ($ \epsilon $) using Eq.~\eqref{fig:ssp-1} leads to the incorrect numrical values. By comparing the values of  $ D $ and $ \epsilon $ calculated from the Eq.~\eqref{zakssp} and Eq.~\eqref{modifiedssp1} (or Eq.~\eqref{modifiedssp2}) from the Table \ref{tab:table1} we have estimated that  the  Eq.~\eqref{fig:ssp-1}  underestimate  crystallite size around $ \simeq $ 35.09 \%  and overestimate the value of strain with same percentage.  

\section{Conclusions}
In conclusion, we have discussed that the equations with physical significance must be dimensionally homogenous, i.e., the dimensions of the right-hand side must be equal to the dimensions of the left-hand side of the equation. We have shown how a dimensionally inhomogeneous equation will lead to the wrong numerical values of parameters. We have proved that the equation reported by the authors for the size strain plot (SSP) does not follow the dimensional homogeneity, leading to incorrect units and numerical values of the estimated parameters. We also noted an error in the crystallite size values reported by the authors. We have suggested the modified equation for the size strain plot (SSP) method, which overcomes all the issues discussed in this short comment and better estimates the materials' crystallite size and strain values.

\section*{Acknowledgments}
The author would like to acknowledge Deepika Shanubhogue and Sharanu for the fruitful discussion. 



\end{document}